\documentclass[aps,twocolumn,prb,groupedaddress,amsmath,amssymb]{revtex4-2}
\usepackage{dcolumn}
\usepackage{bm}
\usepackage{amsmath}
\usepackage{txfonts}
\usepackage[T1]{fontenc}
\usepackage{xspace}
\usepackage{ulem}
\usepackage{comment}
\usepackage{braket}
\ifx\pdfoutput\undefined
\usepackage[dvipdfmx]{graphicx}
\usepackage[dvipdfmx]{hyperref}
\usepackage[dvipdfmx]{color}
\else
\usepackage{graphicx}
\usepackage{hyperref}
\usepackage{color}
\fi

\graphicspath{
{./image/}
}
\begin{document}

\title{Impact of Stealthy Hyperuniform Magnetic Impurity Configurations \\on Bulk Magnetism in a Two-dimensional Heisenberg Model}
\author{Kota Asakura}
\email{asakura@stat.phys.titech.ac.jp}
\author{Kazuki Yamamoto}
\author{Akihisa Koga}
\affiliation{Department of Physics, Institute of Science Tokyo,
  Meguro, Tokyo 152-8551, Japan}

\date{\today}

\begin{abstract}
We investigate an antiferromagnetic quantum Heisenberg model on a square lattice with high-spin magnetic impurities 
to clarify how random and stealthy hyperuniform impurity configurations influence the bulk magnetic properties. 
Stealthy hyperuniform configurations are generated using generalized cost functions that interpolate 
between square-lattice-like and triangular-lattice-like arrangements. 
Using linear spin-wave theory for the mixed-spin model, 
we demonstrate that triangular-lattice-like arrangements yield a larger average staggered magnetization than 
both random and square-lattice-like cases. 
This enhancement originates from sublattice effects: while the square-lattice-like configuration enforces 
nearest-neighbor impurities to occupy opposite sublattices due to its bipartite structure, 
the triangular-lattice-like arrangement allows same-sublattice nearest-neighbor pairs, 
thereby strengthening cooperative magnetic enhancement.
\end{abstract}

\maketitle
\section{Introduction}\label{sec_intro}

Quantum spin systems have been intensively studied since the discovery of high-temperature superconductivity and 
are known to exhibit a wide variety of magnetic phases driven by quantum fluctuations.
In particular, mixed-spin systems have attracted growing interest 
since they host magnetic properties that do not arise 
in uniform-spin models. 
Indeed, existing compounds such as 
magnetite $\rm Fe_3O_4$~\cite{Verway},
spinel ferrites $M \rm Fe_2O_4$ ($M$=Co, Mg, Mn, Zn, etc.)~\cite{MATHEW200751},
double perovskite compounds~\cite{VASALA20151}, 
and mixed-spin chain compounds $R_2\rm BaNiO_5$ ($R$=Nd, Pr)~\cite{Sachan,Zheludev,Zheludev2}
provide representative realizations of such mixed-spin systems.
Previous theoretical studies have largely focused on single-impurity problems~\cite{Kotov98,Kotov981,Hoglund04,Hoglund03,Luscher05,Shinkevich11} or 
idealized periodic systems~\cite{Fujii_1996,Pati_1997,Kolezhuk_1997,Kolezhuk99,Alcaraz_1997,Brehmer_1997,PhysRevB.55.R3336,PhysRevB.57.13610,Koga_1998,PhysRevB.57.R14008,Koga_1999,Sakai99,Koga_2000,Takushima_2000,Yamamoto00,Koga_Nasu_2019}, 
whereas the impact of impurity configurations has received less systematic attention.

 More recently, high-entropy oxides, characterized by the random intermixing of multiple magnetic ions, have attracted broad attention since they exhibit strong compositional disorder and provide a prominent experimental platform for
studying mixed-spin systems~\cite{Rost15,Sarkar21,Witte19,Mao20,Mazza22}.
These compounds have been extensively studied for their magnetic behavior~\cite{Mao19,Meisenheimer17}, reversible energy storage 
capabilities~\cite{Berardan164,Sarkar18,Qiu19}, exceptionally high dielectric constants~\cite{Berardan16}, electrostatic charge transfer~\cite{Rak16}, and catalytic performance~\cite{Chen18}.
Although the magnetic ions in high-entropy oxides are randomly distributed, some of them are known to exhibit antiferromagnetic order~\cite{Zhang19,Pu23}.
These materials highlight the importance of understanding 
how spatially inhomogeneous spin configurations influence magnetic order. 
However, the randomness inherent in high-entropy oxides limits systematic control of spatial correlations.
Thus, it is highly desirable to systematically clarify 
how the spatial configurations of magnetic ions (impurities) affect magnetic properties.

The concept of hyperuniformity~\cite{Torquato03,Torquato18}
provides a useful and systematic framework
for classifying point distributions based on their long-range behavior.
The point configuration is called hyperuniform
if its variance in a large length scale is smaller than a volume law.
The periodic and quasiperiodic point configurations are hyperuniform,
and their order metrics, which measure the degree of the regularity of
the point configuration,
have been examined~\cite{Torquato03,Torquato18,Zachary09,Lin17,Koga24}.
Disordered hyperuniform point patterns have been actively
investigated in recent years and 
are known to appear in a variety of fields such as
soft matter~\cite{Berthier11,Kurita11}, solids~\cite{Zheng16,Llorens20},
active matter~\cite{Huang21}, biology~\cite{Jiao14},
and cosmology~\cite{Gabrielli02}. 
A particularly useful subclass is the stealthy hyperuniform (SHU) system, characterized by the structure factor
$S(\bm k)=0$ for $0<|\bm k|\le K$\cite{Torquato15,Torquato08,Leseur16},
where $K$ is a certain cutoff parameter. 
It remains unclear whether the unique geometric correlations of SHU structures 
can induce local magnetic ordering in mixed-spin systems that is distinct from that in random configurations.
So far, the SHU structures are known to influence physical properties 
in photonic materials~\cite{Florescu09Jan,FroufePere16,Man10,Man13,asakura25}, 
and their impact on quantum spin systems remains largely unexplored~\cite{Bose21}.

In our study, we consider an antiferromagnetic quantum Heisenberg model on a square lattice 
with high-spin magnetic impurities as a minimal theoretical setting.
Using linear spin-wave theory, we take quantum fluctuations into account
and study how their configurations affect the bulk staggered moments.
We find that the average staggered magnetization is strongly affected
by whether nearest-neighbor impurity pairs occupy the same or opposite sublattices,
leading to either enhancement or suppression of the local magnetic moments
depending on their relative positions.
In contrast to random and square-lattice-like SHU arrangements, 
triangular-lattice-like SHU impurity configurations exhibit 
a larger staggered magnetization.

The paper is organized as follows.
In Sec.~\ref{sec_methods},
we first introduce the mixed-spin antiferromagnetic Heisenberg model
on a square lattice, and
explain the spin-wave theory for evaluating magnetic properties.
The procedures for generating SHU point patterns are also given.
In Sec.~\ref{sec_results}, 
we present the numerical results,
analyzing the staggered magnetization for different impurity configurations.
Finally, conclusions are given in Sec.~\ref{sec_conclusion}.

\section{Model and methods}\label{sec_methods}

We consider the two-dimensional antiferromagnetic quantum Heisenberg model
on a square lattice, which is described by the Hamiltonian 
\begin{equation}\label{Heisen}
H = J \sum_{\langle ij \rangle} \bm{S}_i \cdot \bm{S}_j,
\end{equation}
where $\bm{S}_i$ denotes the spin operator at site $i$.
Here, $J$ is the antiferromagnetic exchange coupling constant and
the summation runs over all nearest-neighbor pairs $\langle ij \rangle$.
\begin{figure}[hbt]
  \centering
\includegraphics[width=0.8\linewidth]{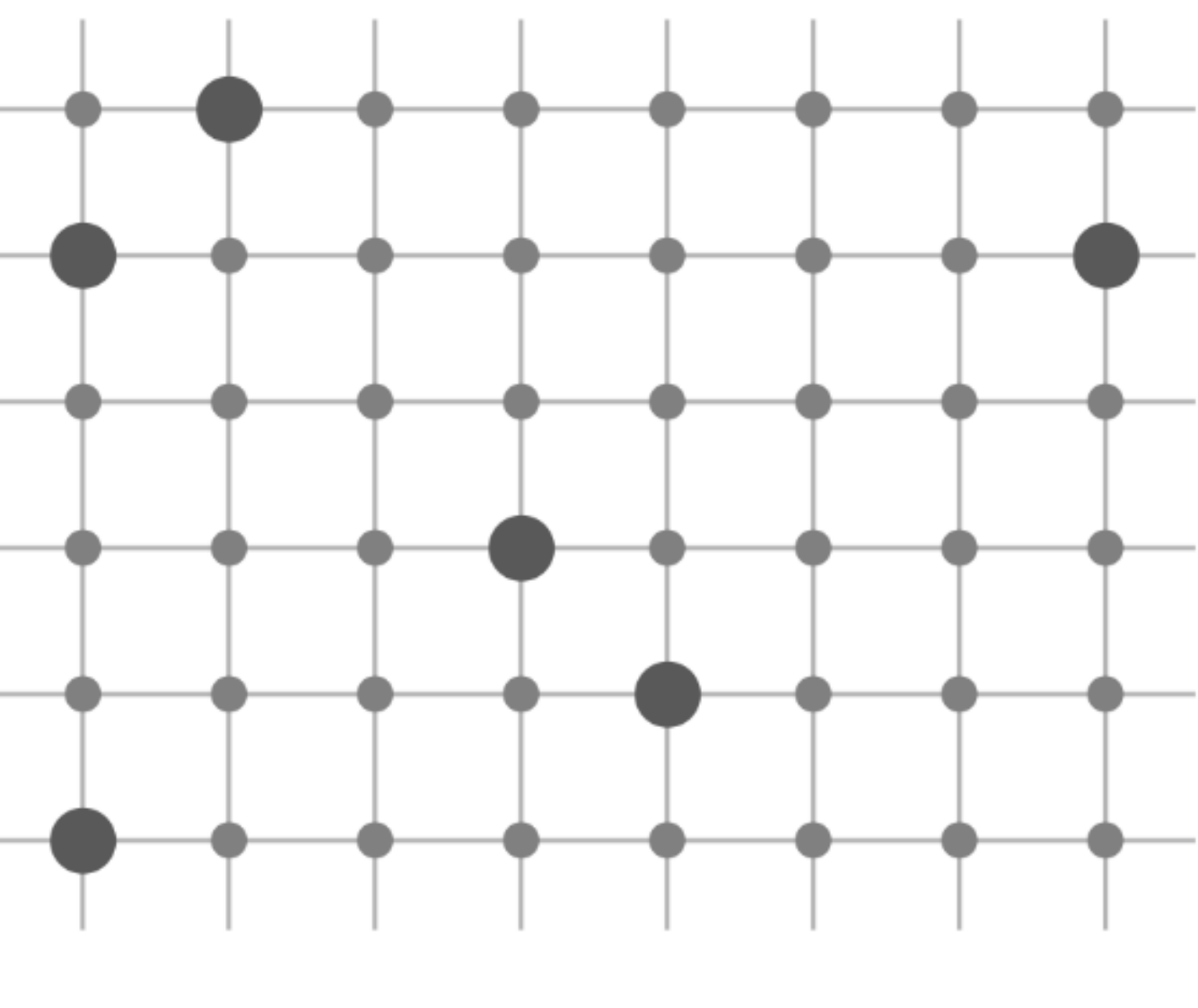}
  \caption{
   Antiferromagnetic quantum Heisenberg model with higher-spin impurities. 
}
  \label{ABAB}
\end{figure}

In this paper, we study the effect of higher-spin magnetic impurities
embedded in a conventional antiferromagnet on a square lattice
with uniform spins $S=1/2$, as shown in Fig.~\ref{ABAB}.
Specifically, we introduce magnetic impurities with spin $S=2$;
that is, the local spin magnitude is taken as $S_i=1/2$ for the host sites,
and $S_i=2$ for the impurity sites.
To ensure that the antiferromagnetically ordered ground state
is realized without a net uniform magnetization,
the magnetic impurities are introduced in equal numbers on each sublattice.
For the numerical calculations,
we consider a spin system with $N=L\times L$ sites as a supercell, and
construct a periodic structure by repeating this supercell,
where $L$ is the even number.
Periodic boundary conditions are imposed with a period $Ln$,
where $n$ is a positive integer representing the number of supercells
along each direction.
In the following, the coordinate of a site is represented as
$\bm{R}_i+\bm{\delta}_m$,
where $\bm{R}_i$ is the coordinate of the $i$th supercell
and $\bm{\delta}_m$ is the relative coordinate of the $m$th site within the supercell.

To examine magnetic properties in the mixed-spin system,
we use linear spin-wave theory based on the $1/S$ expansion.
Specifically, we employ the Holstein-Primakoff transformation,
where the spin operators are expressed for the sublattice $A$,
\begin{align}
S_{im}^{z} &= S_m - a_{im}^\dagger a_{im},\label{spinwave1} \\
S_{im}^{+} &= \sqrt{2S_m - a_{im}^\dagger a_{im}} \; a_{im}\simeq \sqrt{2S_m} \, 
a_{im}, 
\end{align}
and for the sublattice $B$,
\begin{align}
S_{im}^{z} &=- S_m + a_{im}^\dagger a_{im}, \\
S_{im}^{+} &= a_{im}^\dagger \; \sqrt{2S_m - {a_{im}} a_{im}^\dagger} \simeq
\sqrt{2S_m} \, a_{im}^\dagger,\label{spinwave2}
\end{align}
where $a_{im}^\dag (a_{im})$ creates (annihilates) a boson at the $m$th site
in the $i$th supercell.
Furthermore, we use the Fourier transform as
\begin{equation}
  a_{im}=\frac{1}{n}\sum_{\bm{k}}{a_{\bm{k}m}} \exp\Big[ i\epsilon_m \bm{k}\cdot (\bm{R}_i+\bm{\delta}_m) \Big]
    \label{eq_fourier}
\end{equation}
where $\epsilon_m = +1 (-1)$ when the site $m$ belongs to the sublattice $A$ ($B$)
and the wave vector is defined as
$
\bm{k} = (k_x, k_y) = 
\left(2\pi n_x /(Ln), 2\pi n_y /(Ln)\right)$ with 
$n_\alpha = 0, 1, 2, \dots, n - 1$ $(\alpha=x,y)$.
To leading order, the Hamiltonian eq.~\eqref{Heisen} reads
\begin{align}
  H &= \sum_{\bm{k}}
  \begin{pmatrix}
    \bm{a}_k^\dag & \bm{a}_k
  \end{pmatrix}
  \begin{pmatrix}
    h & \Delta_{\bm{k}} \\
    \Delta_{\bm{k}}^\dagger & h
  \end{pmatrix}
  \begin{pmatrix}
    \bm{a}_k\\ \bm{a}_k^\dag
  \end{pmatrix}
\label{Hprime}
\end{align}
where $\bm{a}_k=(a_{k1}\; a_{k2}\; \cdots \; a_{kN})^t$.
Diagonal matrix $h$ is given as
\begin{align}
  h_{mn}=\frac{1}{2}\delta_{mn}\sum_{l\in {\rm NN}(m)}S_l,
\end{align}
where ${\rm NN}(m)$ denotes the set of nearest-neighbor sites of site $m$, and 
\begin{align}
  (\Delta_{\bm{k}})_{mn}=\frac{1}{2}\sum_{\bm{d}\in {\rm NN}(m\rightarrow n)}\sqrt{S_mS_n}e^{ i\epsilon_m \bm{k}\cdot\bm{d}}
\end{align}
where $\bm{d}$ is the displacement vectors from site $m$ to a nearest-neighbor site $n$.

To diagonalize the quadratic bosonic Hamiltonian in Eq.~\eqref{Hprime}, 
we perform a Bogoliubov transformation:
\begin{align}\label{Bogo}
    \begin{pmatrix}
    \bm{a}_k\\ \bm{a}_k^\dag
    \end{pmatrix}
    =
 \begin{pmatrix}
u_{\bm{k}} & v_{\bm{k}}^* \\
v_{\bm{k}} & u_{\bm{k}}^*
 \end{pmatrix}
     \begin{pmatrix}
    \bm{b}_k\\ \bm{b}_k^\dag
    \end{pmatrix},
\end{align}
where $\bm{b}_k=(b_{k1}\; b_{k2}\; \cdots \; b_{kN})^t$ and
$b_{km}$ $(b_{km}^\dag)$ denotes the annihilation (creation) operator
of the quasiparticle, and
$u_{\bm{k}}$ and $v_{\bm{k}}$ are the $N\times N$ matrices.
Details of the diagonalization procedure are given in Appendix~\ref{app}.
Then, we obtain
\begin{align}
  H=\sum_{\bm{k}m} \omega_{\bm{k}m} b_{\bm{k}m}^\dag b_{\bm{k}m},\label{diag}
\end{align}
where $\omega_{\bm{k}m}$ is the quasiparticle excitation energy.

Taking the expectation value with respect to
the ground state of the Hamiltonian
eq.~\eqref{diag}, we obtain the magnetization at site $m$
within a unit cell given by
\begin{align}
\langle S_{im}^z \rangle 
&=\epsilon_m \left[ S_m-\left(\frac{L}{2\pi}\right)^2\int d\bm{k} \sum_n \Big|(v_{\bm{k}})_{mn}\Big|^2\right].
\label{mAB}
\end{align}
Using this local magnetization,
the spatially-averaged staggered magnetization is given as
\begin{align}
M_{\mathrm{AF}}&=\frac{1}{Nn^2}\sum_{im}\epsilon_m\langle  S_{im}^z\rangle.
\end{align}
Diagonalizing the spin Hamiltonian,
we examine bulk magnetic properties of an infinite system constructed
by periodically repeating the supercell.

\subsection{Spatical configurations of magnetic impurities}\label{sec:SHU}

In this study, we discuss how the spatial configurations of the magnetic impurities
affect bulk magnetic properties in the mixed-spin model.
The magnetic impurities in the supercell
are located by $\bm{r}_\ell^{(\gamma)}$ with the sublattice $\gamma$.  
The structure factor is then defined by~\cite{Uche06}
\begin{align}
  S (\bm{k}) &=\frac{1}{N^\prime}\left|\sum_\gamma \rho_\gamma(\bm{k})\,\right|^2,\label{sofk}\\
  \rho_\gamma(\bm{k})&= \sum_{\ell} e^{i \bm{k}\cdot \bm{r}_\ell^{(\gamma)}}. \label{ck}
\end{align}
where $N^\prime$ is the number of magnetic impurities in the supercell, 
and $\rho_\gamma(\bm{k})$ is the Fourier transform
of the coordinates of the magnetic impurities in the sublattice $\gamma$.
When periodic boundary conditions are imposed on the supercell,
the allowed component of $\bm{k}$ is given by
$k_\alpha = 2 \pi n_\alpha/L$ with
integers $n_\alpha (\alpha=x,y)$.

To generate a point pattern with a certain structure factor,
we introduce the objective function $\Phi\:(\ge 0)$, given by
\begin{equation}
  \Phi\left(\{\bm{r}^{(A)}\}, \{\bm{r}^{(B)} \}\right) = \sum_{\bm k} V(\bm k)\Big[S(\bm k) - S_{\mathrm{target}}(\bm k)\Big]^2,
  \label{eq_phi}
\end{equation}
where $S_{\mathrm{target}}(\bm{k})$ is the target structure factor, and 
$V(\bm{k})$ is the window function.
By minimizing the objective function with respect to
$\{\bm{r}^{(A)}\}$ and $\{\bm{r}^{(B)}\}$,
one can obtain the point pattern
corresponding to the target structure factor.

In this study, we focus on the SHU point patterns,
which are characterized by
$S(\bm{k})=0$ for $0<|\bm{k}|\le K$.
The target structure factor and window function are given as
$S_{\mathrm{target}}(\bm k)=0$ and $V_K(\bm k) =\theta(K-|\bm{k}|)$
with $K$ is some positive number and $\theta(x)$ is the Heaviside step function.
It is known that when $K$ increases,
points tend to separate from each other, and
the nearest-neighbor interpoint distance
becomes larger and more uniform.
For sufficiently large $K$,
long-wavelength (long-range) density fluctuations
for the point pattern are suppressed, 
and the resulting configuration approaches
a triangular-lattice-like ordering within the finite supercell,
which is qualitatively similar to the point distribution
obtained in continuous space~\cite{Uche04}.


\begin{figure}[hbt]
  \centering
  \includegraphics[width=\linewidth]{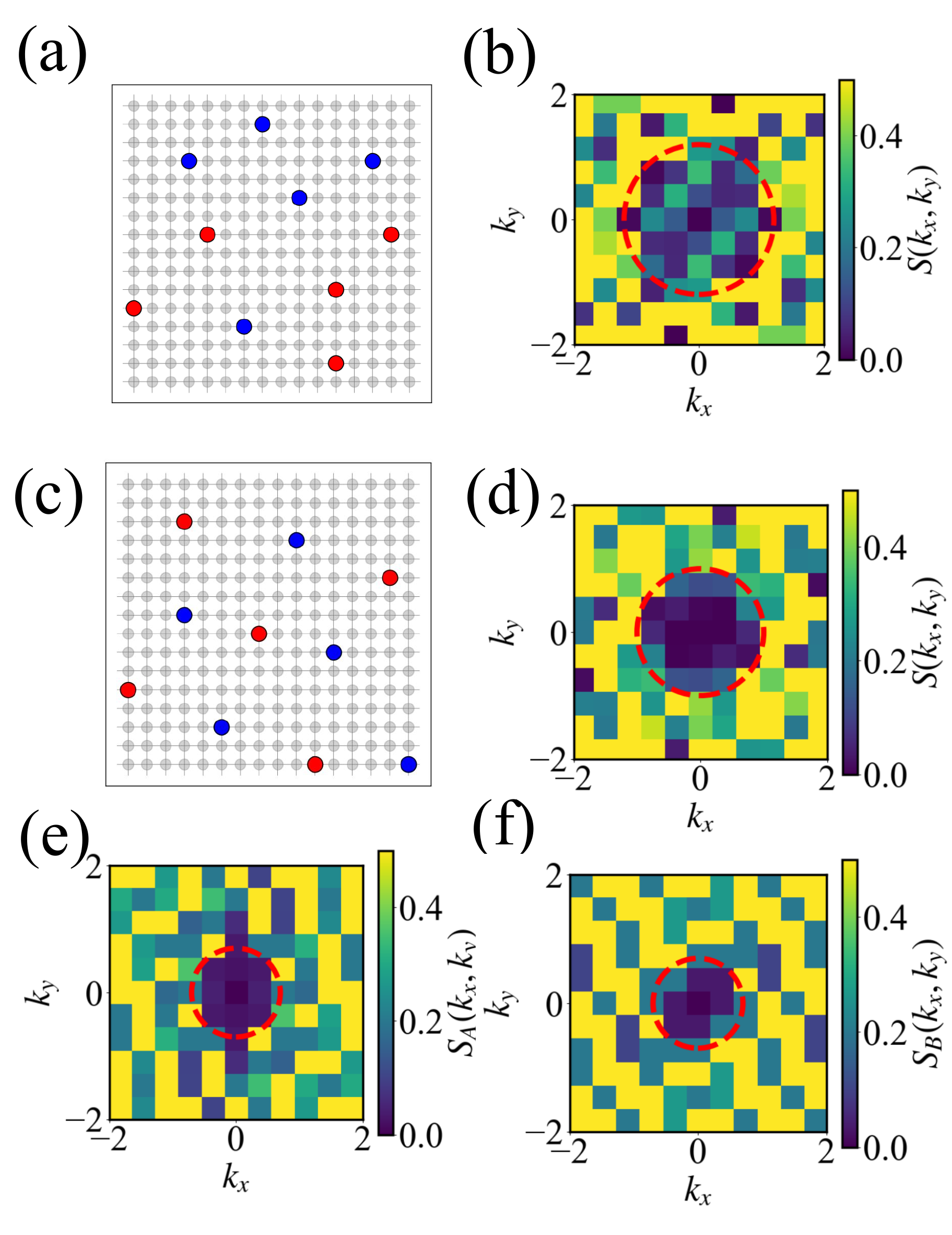}
  \caption{(a) SHU impurity configurations (I). 
    Large red and blue circles represent impurities in the sublattices $A$ and $B$,
    respectively. The gray circles represent $S=1/2$ host spins.
    (b) Structure factor $S(\bm{k})$ of the configuration (I).
    Dashed line indicates the boundary of the window function.
    (c) SHU impurity configurations (II) and its structure factors
    (d) $S({\bm k})$, (e) $S_A({\bm k})$, and (f) $S_B({\bm k})$.
  }
  \label{conf}
\end{figure}
To discuss how the configuration of magnetic impurities 
affects the bulk magnetization,
we also consider an alternative objective function
which confines the magnetic impurities to a single sublattice $\gamma$.
The objective function is defined as
\begin{align}
  &\Phi_\gamma \left(\{\bm{r}_\ell^{(\gamma)}\}\right)
  = \sum_{\bm{k}} V_{K_\gamma}(\bm k)\Big[S_\gamma(\bm k) -S_{\rm target}(\bm{k})\Big]^2,
\end{align}
where $S_\gamma(=2\,|\,\rho_\gamma|\,^2/N^\prime)$ is the structure factor for the magnetic impurities
in the sublattice $\gamma$.
Using this definition, one can consider the total cost function as
\begin{align}
\Psi=(1-c)\Phi +c\sum_\gamma \Phi_\gamma
\end{align}
where $c$ is a constant. 
When $c= 0$ and $K$ is large,
minimizing the cost function drives the magnetic impurities
to separate from each other,
resulting in a triangular-lattice-like spatial configuration.
The impurity configuration for $c=0$ and $K=1.2$
is referred to as the configuration (I).
When $c= 1/2$, and $K$ and $K_\gamma$ are large,
the cost function additionally favors spatial separation among impurities
within the same sublattice.
In this case, the resulting impurity arrangement becomes square-lattice-like.
The impurity configuration for $c=1/2, K=1.1$, and $ K_A=K_B=0.8$ 
is referred to as configuration (II). 
Configurations (I) and (II) are schematically shown in Fig.~\ref{conf}.
This cost function allows us to account for not only
the stealthiness of the magnetic impurities as a whole
but also the stealthiness of impurity distributions within each sublattice.
\begin{figure}[hbt]
  \centering
  \includegraphics[width=\linewidth]{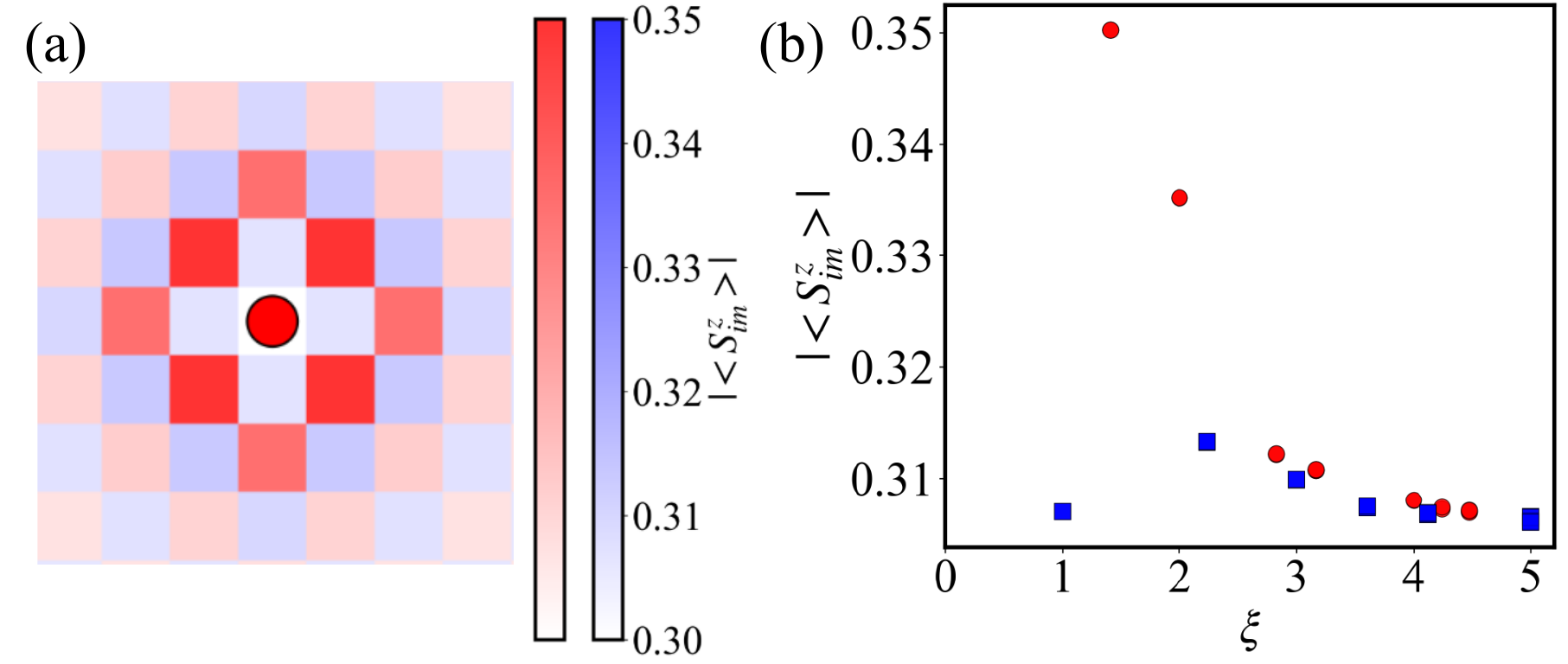}
  \caption{(a) Local magnetization profile around an isolated magnetic impurity. A red circle denotes the $S = 2$ spin.
    The magnitude of the magnetic moment is represented by the density plot, 
    with red (blue) corresponding to the same (opposite) sublattice as the impurity site.
    (b) Squares (circles) represent the magnitudes of the local magnetization
    in the same (opposite) sublattice as an isolated impurity
  as a function of the distance $\xi$ from the impurity site.}
  \label{2imp}
\end{figure}

Before discussing the effects of impurity configurations,
we first examine how a single magnetic impurity affects the magnetic properties 
of the $S=1/2$ quantum Heisenberg model.
To this end, we consider the $S=1/2$ quantum Heisenberg model with $L=16$,
into which two isolated $S=2$ spins are introduced as magnetic impurities.
This allows us to maintain a zero total uniform magnetization
within spin-wave theory while focusing on the local physics of a single impurity.
The local magnetization profile around one of the magnetic impurities
is shown in Fig.~\ref{2imp}(a).
We find that the magnetic moment at the $S=2$ impurity site is renormalized
down to approximately $1.47$, 
which is significantly smaller than the value $\sim 1.8$
obtained for a quantum Heisenberg antiferromagnet with a uniform spin $S=2$. 
This substantial reduction reflects quantum fluctuations originating from
the surrounding $S=1/2$ spin background.
In contrast to the $S=2$ impuruty spins, the moments for the surrounding $S=1/2$ spins
become larger due to quantum fluctuations.
In fact, enhanced magnetic moments are observed in the vicinity of the impurity, 
particularly on nearby sites belonging to the same sublattice  as the impurity site.
In contrast,
when the nearest-neighbor sites of the impurity belong to the opposite sublattice, 
the magnetic moments remain nearly unchanged.
This behavior is clearly illustrated in Fig.~\ref{2imp}(b), 
where the magnetization on the same sublattice exhibits
a strong distance dependence from the impurity, 
whereas that on the opposite sublattice shows only a weak dependence. 
Thus, a higher-spin impurity generates sublattice-dependent magnetic correlations
extending over several lattice spacings \cite{Kotov98,Kotov981,Shinkevich11}.
This observation implies that the relative sublattice positions of multiple impurities 
should play an important role in determining the magnetic properties 
when a finite impurity concentration is introduced.

\section{Configuration dependence in the magnetization profile}
\label{sec_results}

\begin{figure*}[htb]
  \centering
  \includegraphics[width=\linewidth]{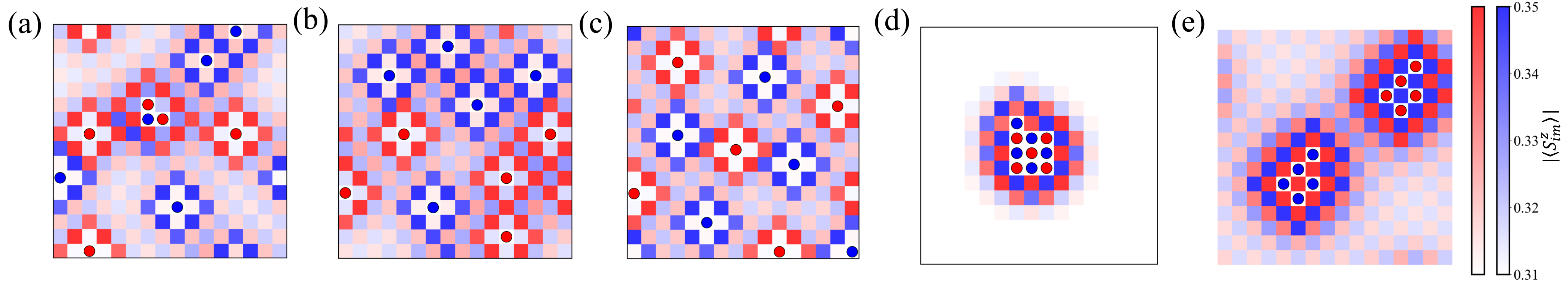}
  \caption{
  Magnetization profiles of the spin system on a $16 \times 16$ square lattice with 10 impurities arranged in (a) a random configuration, (b) configuration (I), and (c) configuration (II). Panels (d) and (e) show the results for systems with clustered impurity configurations. Red (blue) circles denote $S = 2$ spins on sublattice $A$ ($B$).
    The magnitude of the local magnetic moment is represented by the density plot, with red (blue) corresponding to sublattice $A$ ($B$).
  }
  \label{fig2}
\end{figure*}

In this section, we focus on dispersed impurity configurations
in the system with $L=16$, fixing the impurity concentration at $10/256$.
Magnetization profiles for the random and
SHU configurations (I) and (II)
are shown in Figs.~\ref{fig2}(a), (b), and (c).
When the magnetic impurities are spatially well separated,
local magnetic structures similar to those induced by a single impurity
as discussed above
emerge in the vicinity of each impurity.
Namely, sublattice-dependent magnetic correlations
develop locally around individual impurity sites.
We therefore focus on how these local responses
combine to form cooperative magnetic correlations.
In the system with randomly distributed impurities,
both the distances between impurities and
the sublattices of nearest-neighbor impurities depend 
on the local spatial arrangement.
As a result, magnetic correlations are locally enhanced in some regions,
while they are suppressed in others, as shown in Fig.~\ref{fig2}(a).
On the other hand, in the SHU impurity configuration,
the distances between nearest-neighbor impurities are nearly uniform.
In this case, the sublattice arrangement of nearest-neighbor impurities becomes 
the dominant factor controlling the cooperative magnetic response.
In the configuration (I),
the triangular-lattice-like structure appears in the impurity distribution.
In this case, nearest-neighbor impurity pairs sometimes
belong to the same sublattice,
which strongly enhances magnetic correlations between these impurities,
as shown in Fig.~\ref{fig2}(b).
This leads to an overall enhancement of the magnetization.
By contrast, in configuration (II), the magnetic impurities are located on
square-lattice-like sites.
In this case, magnetic correlations are not cooperatively enhanced
since nearest-neighbor impurity pairs belong to distinct sublattices,
as shown in Fig.~\ref{fig2}(c).

The above results for the systems with dispersed impurity configurations
demonstrate that the sublattice structure of nearest-neighbor impurities
plays an important role in enhancing magnetic correlations
throughout the system.
Motivated by this result,
one might expect that the larger staggered magnetization is realized in
the clustered structure.
To examine this expectation,
we consider two clustered configurations with the same impurity concentration.
Local magnetization profiles are shown in Figs.~\ref{fig2}(d) and (e).
When all magnetic impurities are clustered,
enhanced magnetization appears only in the vicinity of the cluster, while
little enhancement is observed away from it, as shown in Fig.~\ref{fig2}(d).
This behavior indicates that the large magnetic moments in the cluster,
composed of $S=2$ spins, tend to partially screen each other
due to quantum fluctuations,
resulting in suppressed magnetic correlations away from the cluster. 
Similar suppressions are clearly observed
when two adjacent magnetic impurities are introduced into the spin system.
In this case, their magnetic moments are strongly reduced $\sim 1.2$,
and little enhancement is observed outside the $S=2$ dimer (not shown). 
This is analogous to the well-known fact that a magnetic dipole produces
only a tiny magnetic field at long distances.
Therefore, this clustered impurity configuration leads to
a small bulk staggered magnetization.

By contrast, in a clustered system where each cluster is composed of impurities
belonging to only one sublattice,
magnetic correlations are cooperatively enhanced both inside and around the clusters,
as shown in Fig.~\ref{fig2}(e).
This leads to a significantly larger bulk staggered magnetization of $M=0.39$.
These results demonstrate that clustering alone is not sufficient to maximize the average staggered magnetization; 
rather, the sublattice-dependent arrangement of impurities is essential for 
achieving cooperative magnetic enhancement across the system.

\begin{figure}[hbt]
  \centering
\includegraphics[width=\linewidth]{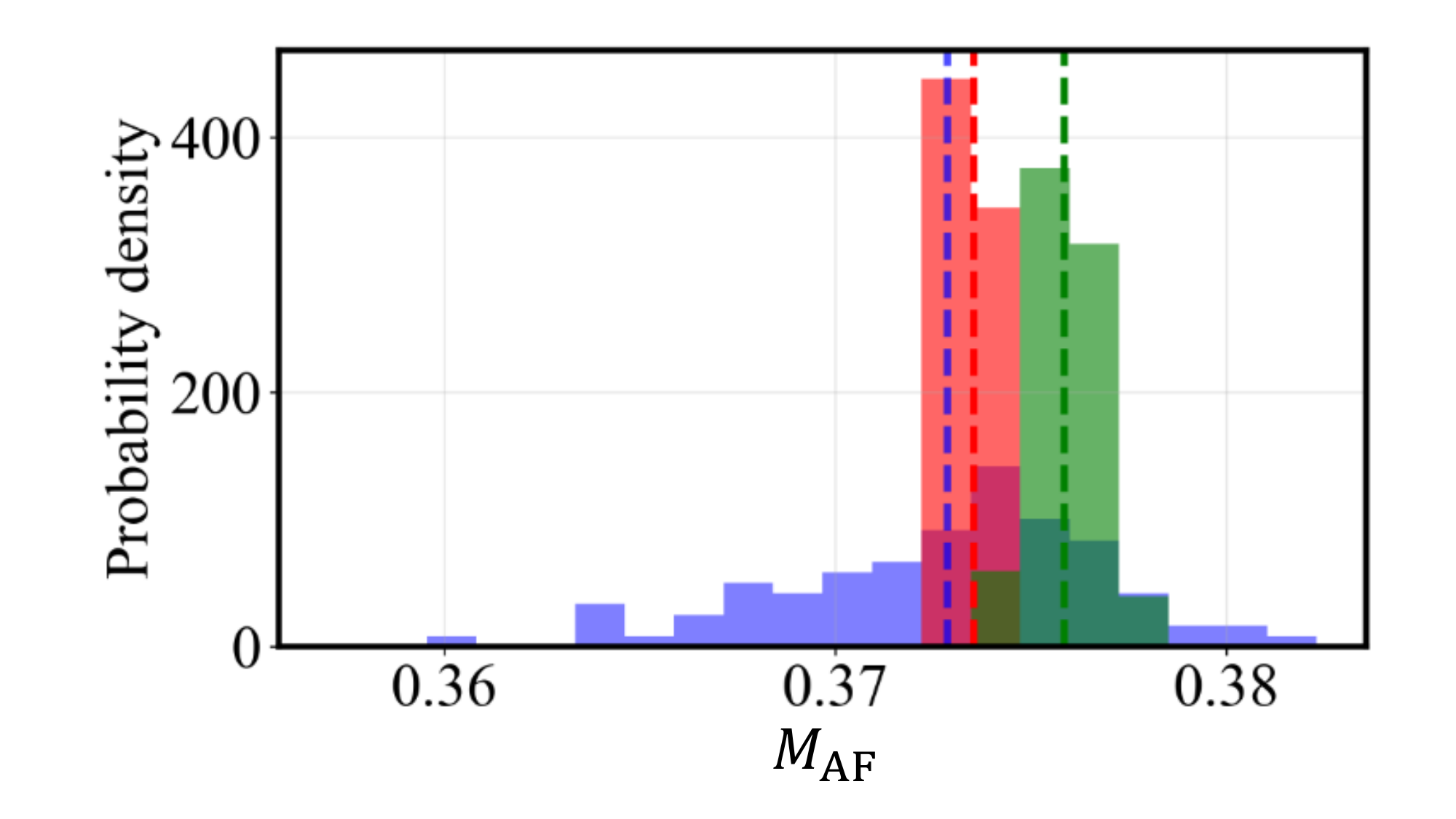}
  \caption{Normalized histgrams of the average sublattice magnetization $M_{\mathrm{AF}}$ for random and SHU impurity configurations. 
    Comparison between random point patterns (blue, 100 samples) and SHU configuration (I) (green, 40 samples). Dashed lines indicate the corresponding mean values. 
    Comparison between random point patterns (blue, 100 samples) and SHU configuration (II) (red, 40 samples). 
}

  \label{fig5}
\end{figure}
Next, we examine the statistical properties of the bulk staggered magnetization
over independent impurity configurations.
Figure~\ref{fig5} shows the histogram of the staggered magnetization
in quantum spin systems with randomly distributed impurities and
SHU configurations (I) and (II).
We find that the staggered magnetization is widely distributed
for the random impurity configurations since
the magnetic properties strongly depend on the specific impurity realization.
The average value is $M_{\mathrm{AF}} = 0.3729$
with a standard deviation $\sigma = 0.0041$.
By contrast, in the SHU configurations,
the distances between nearest-neighbor impurities are nearly identical,
and all samples share similar structural properties.
Configuration (I) nevertheless exhibits a larger sample-averaged staggered magnetization
with $M_{\mathrm{AF}}=0.3759$ and $\sigma=0.001$.
In the system with SHU configuration (II),
where the square-lattice-like impurity arrangement is realized,
the staggered magnetization is much more narrowly distributed,
with $M_{\mathrm{AF}} = 0.3735$ and $\sigma = 0.0004$.
These results can be understood as follows. In the triangular-lattice-like configuration, 
impurities are arranged at nearly uniform distances, 
and nearest-neighbor pairs are not constrained to belong to opposite sublattices. 
This allows same-sublattice impurity pairs to form, leading to a cooperative enhancement of the average staggered magnetization.
In contrast, the square-lattice-like configuration, despite its nearly uniform impurity spacing, 
enforces opposite-sublattice nearest-neighbor pairs due to its bipartite structure. 
Consequently, such cooperative enhancement does not occur.
In the random configuration, short-distance impurity pairs may appear on either the same or opposite sublattices. 
Because these enhancing and suppressing effects occur with comparable probability, 
they statistically compensate each other, 
resulting in an average magnetization similar to that of the square-lattice-like configuration.

We clarify how the average sublattice magnetization correlates with
spatial configurations among $S=2$ impurity spins.
Here, we define a quantity characterizing the spatial configuration of the $S=2$ impurities as
\begin{align}
F\left(\{\bm{r}\}\right)&=
\frac{2}{N_2(N_2-1)}
\sum_{<i,j>} \epsilon_i\epsilon_j
\exp\!\left[ -\frac{1}{2} \left(\frac{r_{ij}}{2}\right)^2 \right],\label{F}
\end{align}
where $\bm{r}_i$ is the coordinate of the $i$th impurity,
$N_2$ is the total number of $S=2$ impurities, and 
$r_{ij}=|\bm{r}_i-\bm{r}_j|$. 
The quantity takes larger values when spins belonging to the same sublattice, 
both with magnitude $S=2$, are located close to each other, 
reflecting strong short-range alignment within the sublattice. 
Conversely, if spins with $S=2$ from different sublattices are close to each other, 
their contributions partially screen out, resulting in smaller values.
Thus, this quantity serves as a measure of the averaged staggered magnetization induced by the impurity configuration.
\begin{figure}[hbt]
  \centering
  \includegraphics[width=\linewidth]{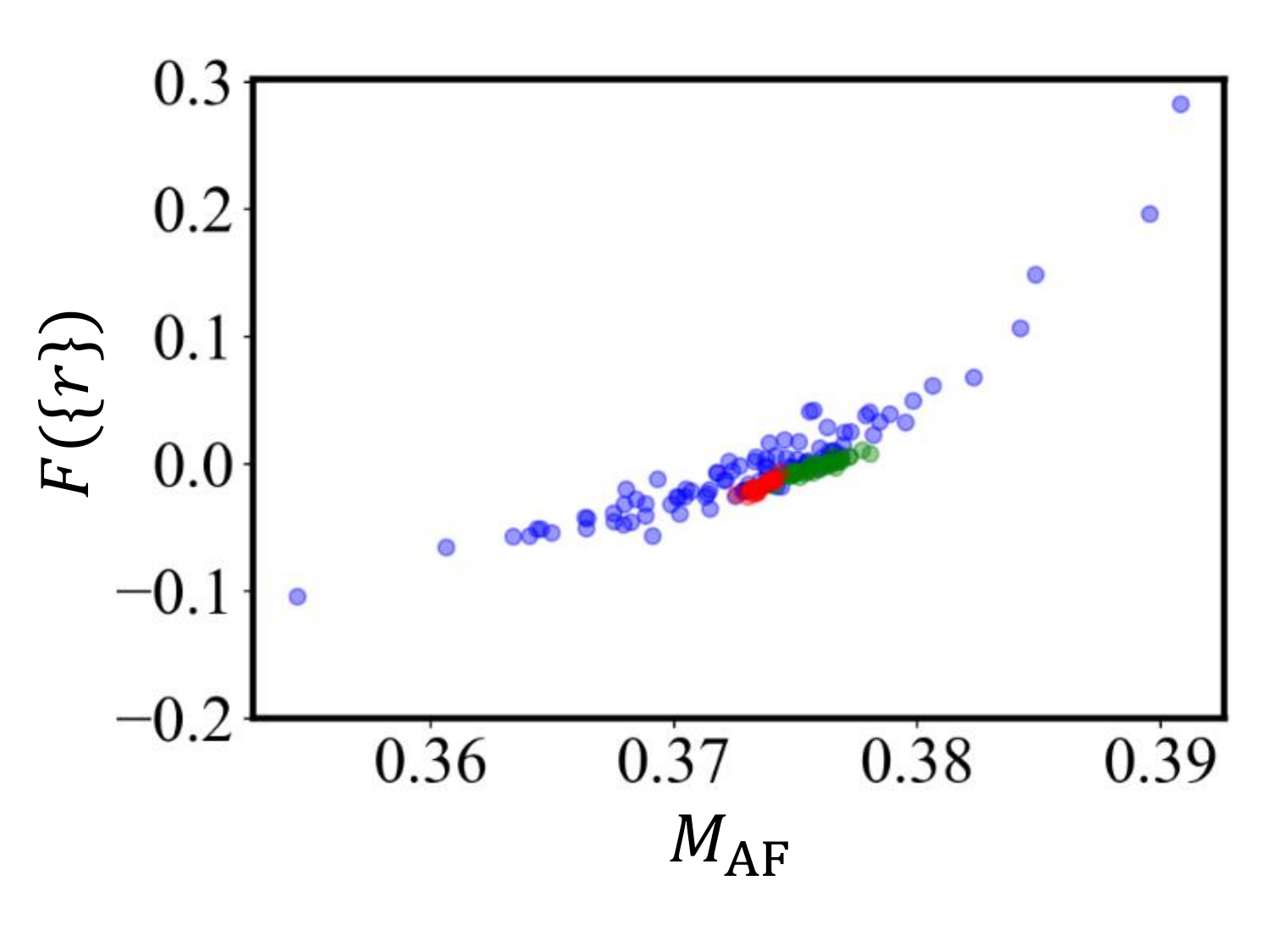}
  \caption{
Relationship between the average sublattice magnetization $M_{\mathrm{AF}}$ 
and the quantity $F(\{\bm{r}\})$
 [see Eq.~\eqref{F}]. 
Blue, green, and red circles represent the results for random configurations, configurations (I) and configuration (II).
}
\label{fig6}
\end{figure}

Figure~\ref{fig6} presents a scatter plot of this quantity obtained from independent samples 
for the random configurations, and the SHU configurations (I) and (II).
First, we find that the values calculated for various impurity distributions
collapse onto a single curve.
This fact indicates that the curve captures
the essential magnetic characteristics of the system.
As $M_{\mathrm{AF}}$ increases, this quantity increases monotonically.
Moreover, it is immediately seen that configuration (I) yields systematically larger average values than configuration (II), reflecting the stronger short-range correlations among same-sublattice impurities in configuration (I).
This behavior is consistent with the fact that the averaged staggered magnetization is enhanced 
when impurities on the same sublattice are located close to each other.
Although the present study is limited to relatively small system sizes
and a single impurity density, 
the observed correlation between $M_{\mathrm{AF}}$ and $F(\{\bm{r}\})$ suggests that 
the quantity provides a useful quantification of local spin correlations and 
can serve as a complementary tool for characterizing the spatial structure of mixed-spin systems.
Overall, our results indicate that the magnetic properties can be systematically characterized from 
the spatial arrangement of magnetic impurities, 
which may serve as a guideline for designing systems with enhanced magnetic moments.

\section{Conclusions}\label{sec_conclusion}
We have investigated the effects of spatial impurity configurations in 
a mixed-spin antiferromagnetic Heisenberg model on a square lattice. 
By systematically comparing random and stealthy hyperuniform impurity arrangements within linear spin-wave theory, 
we demonstrated that stealthy hyperuniform configurations lead to a markedly reduced variance 
in the distribution of the staggered magnetization, 
indicating more uniform magnetic responses across impurity realizations. 
Moreover, we demonstrated that triangular-lattice-like stealthy hyperuniform configurations yield 
a larger average staggered magnetization than both random and square-lattice-like arrangements. 
This enhancement originates from the sublattice dependence of nearest-neighbor magnetic impurities: 
impurities located on the same sublattice cooperatively enhance local magnetic moments, 
whereas those on opposite sublattices suppress them. 
Our results highlight the importance of controlled spatial correlations in impurity configurations and 
demonstrate that hyperuniform structures provide a powerful framework for engineering magnetic properties 
in mixed-spin quantum systems.

\begin{acknowledgments}
  Parts of the numerical calculations were performed
  in the supercomputing systems in ISSP, the University of Tokyo.
  This work was supported by Grant-in-Aid for Scientific Research from
  JSPS, KAKENHI Grants No.\ JP25K17327 (K.Y.),
  JP22K03525, JP25H01521, and JP25H01398 (A.K.).
  This work was partly funded by Hirose Foundation,
  the Precise Measurement Technology Promotion Foundation, and
  the Fujikura Foundation.
\end{acknowledgments}

\appendix
\renewcommand{\thesection}{\Alph{section}} 

\section{Detailed calculation for the Bogoliubov transformation}\label{app}
In this appendix, we give the detailed calculation for choosing the matrix given in Eq.~\eqref{Bogo} in the Bogoliubov transformation.
First, we define
\begin{align}
  H^\prime_{\bm{k}}&=
  \begin{pmatrix}
    h & \Delta_{\bm{k}} \\
    \Delta_{\bm{k}}^\dagger & h
  \end{pmatrix}
\label{Hprime1}
\end{align}
\begin{align}
\tilde{H}_{\bm{k}} &= H^\prime_{\bm{k}}P=
\begin{pmatrix}
h & -\Delta_{\bm{k}} \\
\Delta_{\bm{k}}^\dagger & -h
\end{pmatrix}.\label{HprimeP}
\end{align}
where $H_{\bm{k}}^\prime$ is the matrix given in Eq.~\eqref{Hprime},
and
\begin{align}
P = \mathrm{diag}(\underbrace{1, \dots, 1}_{N}, \underbrace{-1, \dots, -1}_{N}).
\end{align}
We consider the following eigenvalue problem, which is necessary to diagonalize Eq.~\eqref{HprimeP}:
\begin{align}
{\tilde{H}}_{\bm{k}} \Psi_{\bm{k}m} &= \frac{\omega_{\bm{k}m}}{2} \Psi_{\bm{k}m}.
\end{align}
Here, we introduce a generalized inner product, which allows us to diagonalize the equation given in Eq.~\eqref{HprimeP}. 
This generalized inner product is necessary because the standard inner product does not always guarantee the orthogonality or normalization of the eigenvectors:
\begin{align}
\Psi_{\bm{k}m} \circ \Psi_{\bm{k},m^\prime} &\equiv \Psi^{\dagger}_{\bm{k}m} P \Psi_{\bm{k}m^\prime},\label{innerproduct}
\end{align}
which satisfies
\begin{align}
\begin{cases}
(\Psi_{\bm{k}m} \circ \Psi_{\bm{k},m^\prime})^\dagger &= \Psi_{\bm{k}m^\prime} \circ \Psi_{\bm{k}m} ,\\
\Psi_{\bm{k}m} \circ ({\tilde{H}}_{\bm{k}} \Psi_{\bm{k}m^\prime}) &= ({\tilde{H}}_{\bm{k}} \Psi_{\bm{k}m}) \circ \Psi_{\bm{k}m^\prime}.
\end{cases}
\end{align}
Since the generalized inner product defined in Eq.~\eqref{innerproduct} does not fix the magnitude of the vectors, proper normalization and orthogonality condition must be imposed as follows:
\begin{equation}
\Psi_{\bm{k}m} \circ \Psi_{\bm{k}m} \equiv 
\begin{cases}
\delta_{{m},{m^\prime}}, & \text{if }\omega_{\bm{k},m}>0,\\
-\delta_{{m},{m^\prime}}, & \text{if } \omega_{\bm{k},m}<0.
\end{cases}\label{normalization}
\end{equation}
Regarding the eigenvalues $\omega_{\bm{k}m}$, the Hamiltonian in Eq.~\eqref{HprimeP} always has real eigenvalues that are either positive or negative. 
To demonstrate this, we introduce the operator $C$ to represent complex conjugation and define the block matrix $M$ as
\begin{equation}
M = 
\begin{pmatrix}
0 & E \\
E & 0
\end{pmatrix},
\end{equation}
where $E$ is the identity matrix. Then, we have
\begin{equation}
{\tilde{H}}_{\bm{k}}(CM\Psi_{\bm{k}m})
=-CM{\tilde{H}}_{\bm{k}} \Psi_{\bm{k}m}
=-\frac{\omega_{\bm{k}m}}{2} (CM\Psi_{\bm{k}m}).
\end{equation}
This shows that $CM \Psi_{\bm{k},m}$ is also an eigenvector of ${\tilde{H}}_{\bm{k}}$ with eigenvalue $-\omega_{\bm{k}m}/2$, demonstrating the symmetric structure of the eigenvalues.
We also define the set of eigenvectors of $\tilde{H}$ as ${V}_{\bm k}$, which is given by
\begin{align}
{V}_{\bm{k}} &= 
\Bigl( \Psi_{\bm{k},1}, \dots ,\Psi_{\bm{k}m}, \dots,\Psi_{\bm{k}N}, \notag\\
& \quad \quad
CM \Psi_{\bm{k},1}, \dots ,CM\Psi_{\bm{k}m}, \dots, CM \Psi_{\bm{k}N} \Bigr).\label{CM}
\end{align}
which is a $2N \times 2N$ matrix.
By using Eq.~\eqref{normalization}, $V_{\bm{k}}$ satisfies the following generalized normalization and orthogonality condition
\begin{equation}
{V}_{\bm{k}}^\dagger P {V}_{\bm{k}} = P.\label{nuPnu}
\end{equation}
In order to perform the final step of the diagonalization procedure, 
we introduce the relation between ${V}_{\bm{k}}$ and $\Omega_{\bm{k}}$, 
where $\Omega_{\bm{k}}$ is the diagonal matrix of eigenvalues obtained 
from the final diagonalization of $H^\prime_{\bm{k}}$ in Eq.~\eqref{Hprime1}:
\begin{align}
\Omega_{\bm{k}}
&=
\mathrm{diag}
\left(
\underbrace{\omega_{\bm{k},1}, \dots, \omega_{\bm{k},N}}_{N},
\underbrace{\omega_{\bm{k},1}, \dots, \omega_{\bm{k},N}}_{N}
\right).
\end{align}
The relation between ${V}_{\bm{k}}$ and $\Omega_{\bm{k}}$ is then given by
\begin{align}
V_{\bm{k}} \Omega_{\bm{k}} P &= \tilde{H}_{\bm{k}} V_{\bm{k}} .
\label{nuome}
\end{align}
Since we also require an explicit matrix form of the Bogoliubov transformation matrix  in Eq.~\eqref{Bogo}, we choose the matrix $U_{\bm{k}}$ as
\begin{align}
{U}_{\bm{k}}&=P V_{\bm{k}} P.\label{PnuP}
\end{align}
We can determine the form of $U_{\bm{k}}$ based on Eq.~\eqref{CM}:
\begin{align}
U_{\bm{k}} =
\begin{pmatrix}
u_{\bm{k}} & v_{\bm{k}}^* \\
v_{\bm{k}} & u_{\bm{k}}^*
\end{pmatrix}.
\end{align}
For convenience, we rewrite the vectors in Eq.~\eqref{Bogo} as
\begin{align}
    {\bm{\Lambda}}_{\bm{k}}&=
    \begin{pmatrix}
    \bm{a}_k\\ \bm{a}_k^\dag
    \end{pmatrix},
    \\
    {\bm{\Gamma}}_{\bm{k}}&=\begin{pmatrix}
    \bm{b}_k\\ \bm{b}_k^\dag.
    \end{pmatrix}
\end{align}
where $\bm{a}_{\bm{k}}$ and $\bm{b}_{\bm{k}}$ are bosonic operators,
so $U_{\bm{k}}$ must satisfy a condition that preserves the bosonic commutation relations of ${\bm{\Lambda}}_{\bm{k}}$ and $\bm{\Gamma}_{\bm{k}}$. This condition is expressed as
\begin{align}
U_{\bm{k}}^\dagger P U_{\bm{k}} =U_{\bm{k}} P U_{\bm{k}}^\dagger = P.\label{PuP}
\end{align}
This choice of the form of $U_{\bm{k}}$ in Eq.~\eqref{PnuP} is made specifically to ensure that Eq.~\eqref{PuP} is automatically satisfied.
By substituting Eqs.~\eqref{Bogo}, \eqref{HprimeP}, and 
\eqref{nuPnu}-\eqref{PuP} to Eq.~\eqref{Hprime}, we can diagonalize the Hamiltonian $H_{\bm k}^\prime$ as
\begin{align}
    \bm{\Lambda}_{\bm{k}}^\dagger H_{\bm{k}}^\prime \bm{\Lambda}_{\bm{k}}&=
    \bm{\Gamma}_{\bm{k}}^\dagger {U}_{\bm{k}}^\dagger H^\prime_{\bm{k}}  {U}_{\bm{k}} \bm{\Gamma}_{\bm{k}}\notag\\
    &= \bm{\Gamma}_{\bm{k}}^\dagger P {V}_{\bm{k}}^\dagger P \tilde{H}_{\bm{k}} P  P {V}_{\bm{k}} P \bm{\Gamma}_{\bm{k}} \notag\\
    &= \bm{\Gamma}_{\bm{k}}^\dagger \Omega_{\bm{k}} \bm{\Gamma}_{\bm{k}},\label{GOG}
\end{align}
By using Eq.~\eqref{PnuP}, we can derive the final diagonalization relation Eq.~\eqref{GOG} and also compute Eq.~\eqref{mAB} in numerical calculations.

\nocite{apsrev42Control}
\bibliographystyle{apsrev4-2}
\bibliography{example.bib}
\end{document}